# Independent Experimentation on the Activation of Deuterium-Loaded Materials by X-Ray Exposure

Rob Davies
Jet Propulsion Laboratory
Pasadena, CA 91105

April 13, 2016

# Introduction

In an attempt to replicate work performed elsewhere [1], we have searched for x-ray induced nuclear activation of deuterated materials. Our first results, reported in September 2015, showed no evidence of nuclear activation, contrary to the results reported in [1]. The primary shortcoming of our first attempt, however, was that the x-ray tube we used was capable of a maximum accelerating potential of only 160kV, less than the 200kV used on the comparison test sample from [1]. Thus, our results could not exclude processes initiated by x-rays near 200 keV in energy. A second potential shortcoming was that our x-ray tube did not have a microfocus beam, like the one used in [1].

Recently, we have irradiated additional samples with an x-ray tube matching the specifications of the tube used in [1]. By matching the test conditions with essentially identical equipment, we have overcome both of the shortcomings of our previous measurements. As before, we find no evidence of activation.

The bulk of this report, pages 2 through 35, contain the results and analysis from the first round of testing. The extension to 200kV is found in the final appendix, namely Appendix A5.

# Summary

We have explored x-ray induced nuclear activation of deuterated materials. A mixture of deuterated polyethylene and titanium deuteride was exposed to <160 keV x-rays from a tungsten target. Following the exposure, a low-background alpha/beta counter was used to inspect the material for resulting activity. All activity measurements (decays per minute) are zero within uncertainties and are below the detector's Minimum Detectable Activity.

Following a very similar process, previous research [1] had detected both alpha and beta decays. Results for the present work are summarized in Table 1, alongside results for one of the most active samples reported in [1]. For mechanisms that can be initiated with photons up to 144keV our run conditions were more severe than those in [1], in which case our results suggest that the alpha/beta activity observed in [1] were not x-ray induced. If there is a process for alpha/beta activation that depends on exceeding an energy threshold above 144keV then the present work may not have been sensitive to the mechanism.

Given our null result, we propose an alternative theory for the non-zero activities measured previously. Namely, that the materials in the previous work were contaminated with daughters of radon decay, which contain both alpha and beta emitters. The plausibility of this theory is enhanced by the fact that the previous work was conducted partially in a basement laboratory in a part of the country where radon gas is common. Radioactive daughters are electrostatically charged following radon decay, and thus can accumulate on electrostatically charged materials. Pursuant to this theory, we have shown that our PE and DPE materials are easily charged and can retain such a charge for an extended period of time. They are easily discharged with an ion blower, but are not easily discharged through contact with a grounded conductor, which is understandable given that they are electrical insulators. One contraindication of this theory is the absence of a signal in the control samples in the previous study. Both the PE and DPE materials, if electrostatically charged, would be expected to attract radon daughters. This difference could be explained by a systematic difference in method or timing of the processing of the PE and DPE samples used in the previous work. We have no means of exploring whether there was such a difference, and therefore, cannot come to a firm conclusion about the validity of our alternate theory.

**Table 1.** Results summary with comparison to "SL16", a similar test sample from [1].

| Sample | Ref. [1] | Present Work | | | | | | | | Units |
|---|---|---|---|---|---|---|---|---|---|---|
| | SL16 | 1 | 2 | 3 | 4 | 5 | 6 | 7 | 8 | |
| **Materials** | Deuterated | Deuterated | Control | Deuterated | Control | Deuterated | Control | Deuterated | Control | |
| **Mass (TiD: DPE or TiH: PE)** | 0.594: 0.454 | 0.5: 0.5 | 0.5: 0.5 | 0.5: 0.5 | 0.5: 0.5 | 0.5: 0.5 | 0.5: 0.5 | 0.5: 0.5 | 0.5: 0.5 | g |
| **Max X-Ray Energy** | 200 | 160 | 160 | 160 | 160 | 160 | 160 | 160 | 160 | keV |
| **Shielding (Stainless Steel or Aluminum)** | 0.25" SS | 0.14" SS | 0.14" SS | 0.14" SS | 0.14" SS | 0.001" Al | 0.001" Al | 0.001" Al | 0.001" Al | inches |
| **X-Ray Current** | 1 | 1 | 1 | 1 | 1 | 1 | 1 | 2 | 2 | mA |
| **Run Time** | 60 | 90 | 90 | 90 | 90 | 90 | 90 | 90 | 90 | minutes |
| **X-Rays Scattered In Target Mat'l** | 2.95E+13 | 5.75E+13 | 5.75E+13 | 5.75E+13 | 5.75E+13 | 1.43E+15 | 1.43E+15 | 2.86E+15 | 2.86E+15 | count |
| **X-Rays Scattered In Target Mat'l** | 1x | 2x | 2x | 2x | 2x | 25x | 25x | 50x | 50x | ratio to SL16 |
| **Lag Time (x-ray to detection)** | 30 | 10 | 18 | 15 | 21 | 16 | 14 | 13 | 21 | min |
| **Activity After X-Ray - Alpha** | 8.44±2.1 | -0.09±0.09 | 0.43±0.38 | 0.17±0.27 | 0.17±0.27 | -0.09±0.09 | 0.43±0.38 | 0.17±0.27 | 0.95±0.53 | dpm |
| **Min Detectable Activity - Alpha** | 2.54 | 1.27 | 1.27 | 1.27 | 1.27 | 1.27 | 1.27 | 1.27 | 1.27 | dpm |
| **Activity After X-Ray - Beta** | 37.95±6.47 | -0.42±0.92 | 2.09±1.17 | -1.68±0.77 | 0.00±0.97 | -1.47±0.79 | -0.84±0.87 | -0.84±0.87 | 0.84±1.05 | dpm |
| **Min Detectable Activity - Beta** | 6.93 | 3.75 | 3.75 | 3.75 | 3.75 | 3.75 | 3.75 | 3.75 | 3.75 | dpm |

# 1. Introduction

Results from [1] suggest that <200keV x-rays from a tungsten target may activate mixtures of deuterated-polyethylene (DPE) and titanium deuteride (TiD). The reported activity included a relatively short-lived alpha component (persisting for >1/2 hour) and a beta component that remained present >4 months after the x-ray exposure. While nuclear activation by photons at this low of an energy, if confirmed, would be a surprising result, the fact that control samples comprised of hydrogenated versions of the same material (PE and TiH) showed no activation obviates a quick dismissal of the result.

The present work attempts to reproduce the basic results of the previous experiment. For the sake of expediency, we use only one combination of materials, corresponding to one of the most activated samples from the earlier work. In most other ways, we have attempted to duplicate the earlier experiment. For example, we use tungsten target x-rays with a beryllium window, comparable exposure durations, and the same model detector for alpha/beta activity. One difference, however, is that our x-ray source is capable of only 160keV. At this accelerating potential, the majority of the energy spectrum from the previous work is covered, namely the characteristic x-rays and the majority of the Bremsstrahlung energies. To partially compensate we thinned the material between the x-ray tube and the target material and ran longer exposures, resulting in substantially greater integrated fluence.

# 2. Test Setup
## 2.1. Material Description

We procured polyethylene (PE) and deuterated polyethylene (DPE) from Polymer Source of Montreal, Canada[†]. The manufacture's analyses of the PE are found in the Appendix A1. Also found in the same appendix is the manufacturer's Certificate of Analysis for the DPE, which indicates greater than 98% purity. JPL's Analytical Chemistry Laboratory confirmed that the Z=1 atoms in the DPE was almost exclusively deuterium. They measured a C-H to C-D ratio of less than 1%. See Appendix A2.

The PE and DPE are visibly distinct. The PE is greyish white in color and is lower density than the DPE. The DPE is also white, but with a yellow tint. As received both materials are clumped, similar to what had been reported by the previous researchers. See Figure 1.

---

[†] This is one of multiple sources for PE and DPE recommended by the authors of [1].

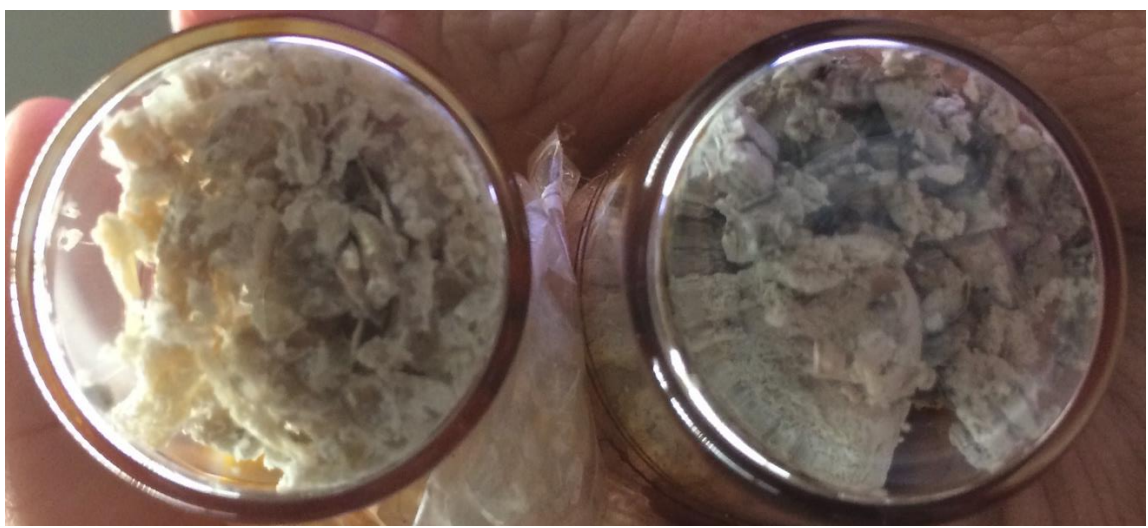

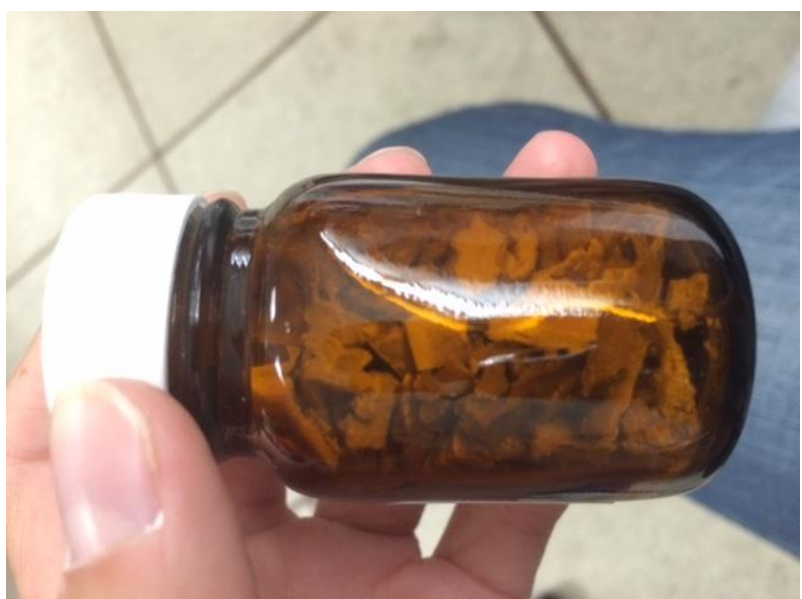

**Figure 1**. Samples of DPE, upper left, and PE, upper right, and a bottle of PE showing clumps, below.

The titanium hydride (TiH) and titanium deuteride (TiD) were procured from Hydrogen Components Incorporated of Bailey, Colorado‡. Both TiH and TiD were ground and filtered to granules sizes between 300um and 1000um (-18+50 mesh). Bottles of the deuterated product are shown in Figure 2. Also, received were the finest particles of material, that which passed through the fine mesh (-50 mesh; <300um). We chose to use the fine material to minimize self-shielding in the sample, as discussed below.

---

‡ This is the source of TiH and TiD used in [1].

The TiH and TiD production process infuses hydrogen or deuterium into the titanium under high pressure. This process leads to a non-stoichiometric ratio of Z=1 atoms to titanium. The production goal for our order was to get close to a 2:1 ratio (i.e. TiD$_2$). The manufacturer determines the ratio during production by weighing the titanium sample before and after the infusion process. For our lot, manufacturer states that the ratio is 1.97:1 (i.e. TiD$_{1.97}$ and TiH$_{1.97}$). No attempt was made to independently verify these atomic ratios.

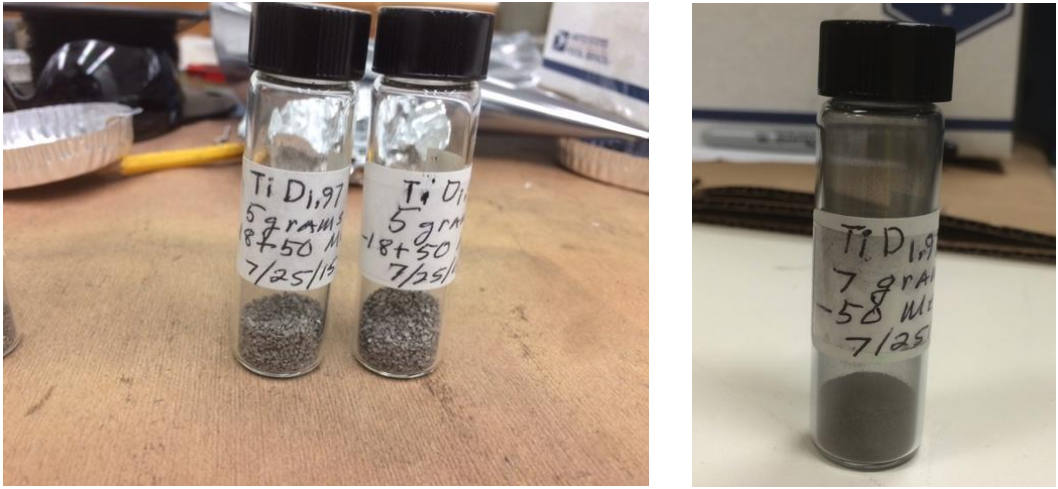

**Figure 2.** Bottles of titanium deuteride.

## 2.2. Sample Preparation

Each of the 4 deuterated test samples consisted of 0.5g of the TiD and 0.5g of DPE. Similarly, each of the 4 control samples consisted of 0.5g of TiH and 0.5g of PE. These ratios were chosen to be comparable to sample SL16, which showed the greatest alpha activity and among the greatest beta activity in [1].

A significant shortcoming of the previous work was the lack of repeatable activity measurements. Repeatable activity measurements are a prerequisite to estimating half-lives and other time dependent effects (e.g. growth of daughter species.) Such measurements could provide important clues to the identity of the active isotopes. As acknowledged by the previous researchers, the lack of repeatability significantly limits the value of their Figure 6 (page 12 of [1]).

The activity measurements were not repeatable because the relative orientation and layering of the sample constituents were different each time the activity of a sample was measured. The orientation of the sample materials mattered all the more because of significant amount of shielding that could be caused by the substantial size of the PE clumps, which would substantially block any alpha or beta particles emitted.

In the present work, two improvements to the experimental technique were employed in an effort to improve repeatability of the activity measurements. First, small pieces of the PE materials were cut from the clumps. Both cutting with a razor and chopping with a high-speed rotary blade were used. The rotary cutter is reminiscent of a small food processor. The glass bowl, metal blades and plastic lid were cleaned with IPA before use.

The second innovation was to line the bottom of the planchet with double-faced tape to stabilize the sample. One planchet was dedicated to each sample. The two pictures in Figure 3 are of the same sample before and after the sample had been tested in the Canberra. Note the stability of the sample as evidenced by the relative positions and orientations of identifiable constituent pieces.

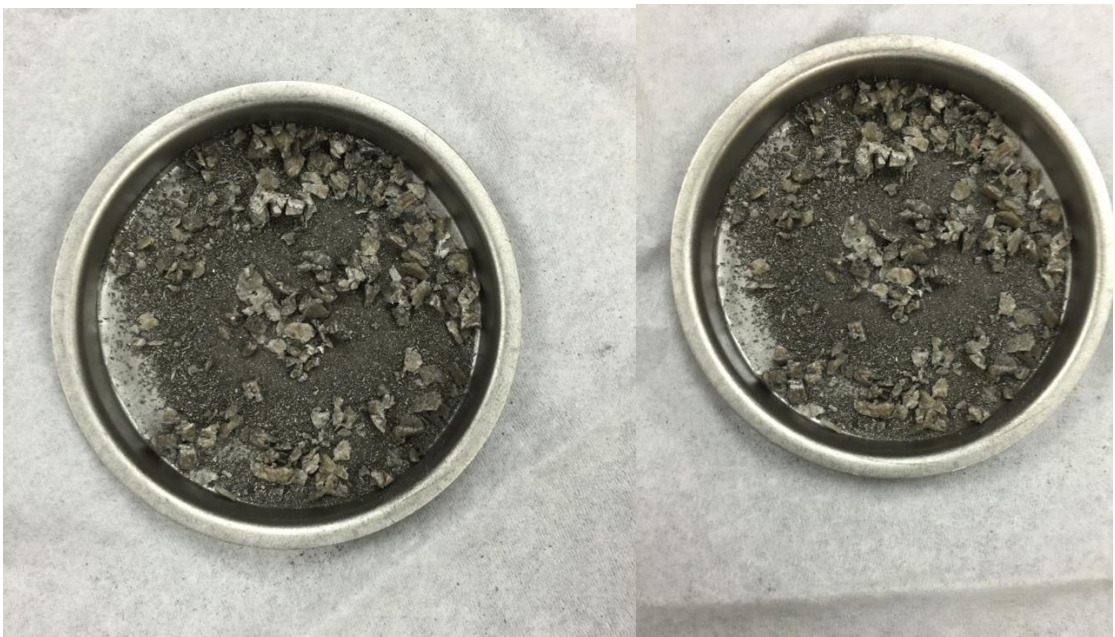

**Figure 3.** Sample 1, in a planchet with double face tape on the bottom, is shown before and after measurements in the Canberra system. Note the stability of the sample as evidenced by the relative positions and orientations of identifiable constituent pieces.

0.5g each of the chopped DPE (PE) and TiD (TiH) were mixed together and poured§ into the test fixture, which is shown in Figure 4. For samples 1 through 4, the fixture was capped off and evacuated with a rough pump. However, for Sample 2 one of the fixture's couplings was discovered later to have a loose nut; thus, it is almost certain that the vacuum seal did not hold for Sample 2. Samples 5 through 8 were irradiated with a layer of aluminum foil in place of the "cap" (see Figure 4), so these samples were not evacuated.

---

§ The density of PE was lower than DPE, so some tamping was required to get it in.

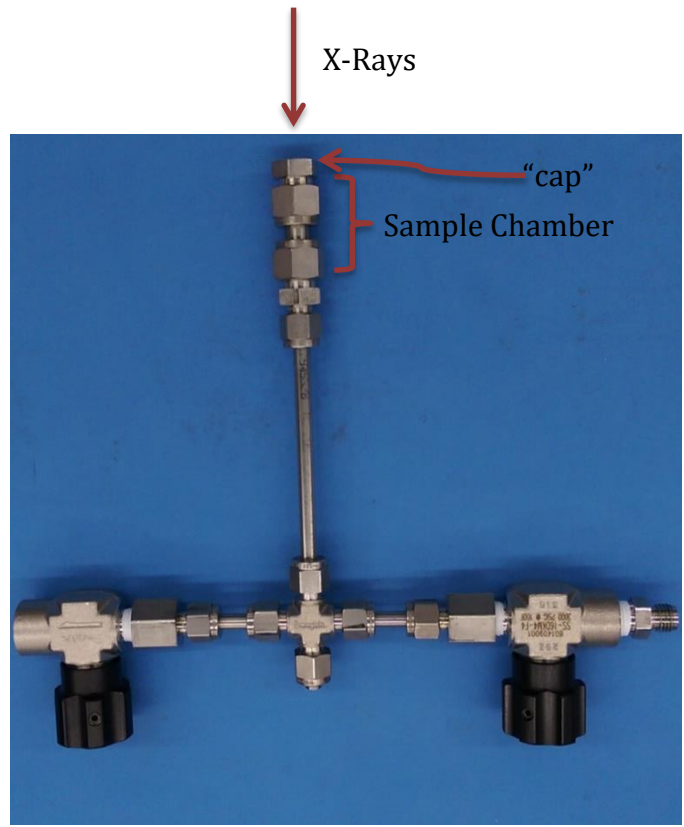

**Figure 4.** Test fixture. The material sample is held in the section shown at the top of the picture.

In addition to closing off the test fixture, the cap also acts as a shielding between the x-ray source and the target material. The cap used in [1] presented 0.25" of stainless steel shielding. To increase the flux in the target materials, we replaced the original cap with one that had been machined down to 0.14". This configuration was used for Samples 1-4. Following the null result for those samples, we further increased the flux by replacing the cap with aluminum foil, 0.001" thick.

## 2.3. Test Equipment

Table 2 provides a summary of the equipment used in the experiment. The two pieces of equipment most central to the investigation are the LoRad x-ray machine and the Canberra alpha and beta counter. The radiation survey meters were used primarily to ensure that the experimenters were not being exposed to significant activity.

Table 2. Equipment summary, including serial number and calibration date.

| Equipment | Manuf'r | Model | SN | Calibration Date |
|---|---|---|---|---|
| X-Ray Machine | LoRad | LPX160 | 102675995-A00031 20/15 | Conformance Inspection 5/12/2015 |
| Canberra Low Alpha/Beta Counter | Canberra | Tennelec S5XLB | 10026919 | 7/9/15 |
| Alpha Scintillator | Ludlum | 3 | 162617 | 5/12/15 |
| Probe for Alpha Scintillator | Ludlum | 43-5 | 167297 | 5/12/15 |
| Survey Meter | Victoreen | 451B | 1073 | 7/14/15 |
| Geiger Counter Survey Meter | LudLum | 4 | 161877 | 2/9/15 |
| Probe for Geiger Counter | LudLum | 44-40 | 165856 | 2/9/15 |
| Electrostatic Fieldmeter (ID IS028529) | Simco | FMX-003 | RX02558 | 9/30/14 |

## 2.4. X-Ray Machine

The capabilities of the LPX160 x-ray machine are summarized in the datasheet found in Appendix A3. The machine can operate with an accelerating voltage up to 160keV with a 5mA beam current. For the present work, it was operated at maximum voltage and either 1 or 2mA, as detailed in Section 3.

In the next three figures, JPL's unit is shown being readied for use. Figure 5 shows the body of the x-ray machine mounted in its tripod. In this image, the body is rotated to show the beryllium window through which x-rays emerge. During our experimentation the body was rotated to direct the x-ray's down toward the floor.

The positioning capability of the tripod was used to position the x-ray source relative to the test sample. Figure 6 shows the test fixture held in place just below the x-ray window. The fixture is held in place using a vertical lab stand and a clamp. In the figure, positioning is not yet complete because the body of the x-ray machine must still be lowered to minimize the distance between the beryllium window and the cap of the test fixture. The final position, after lowering, is shown in Figure 7.

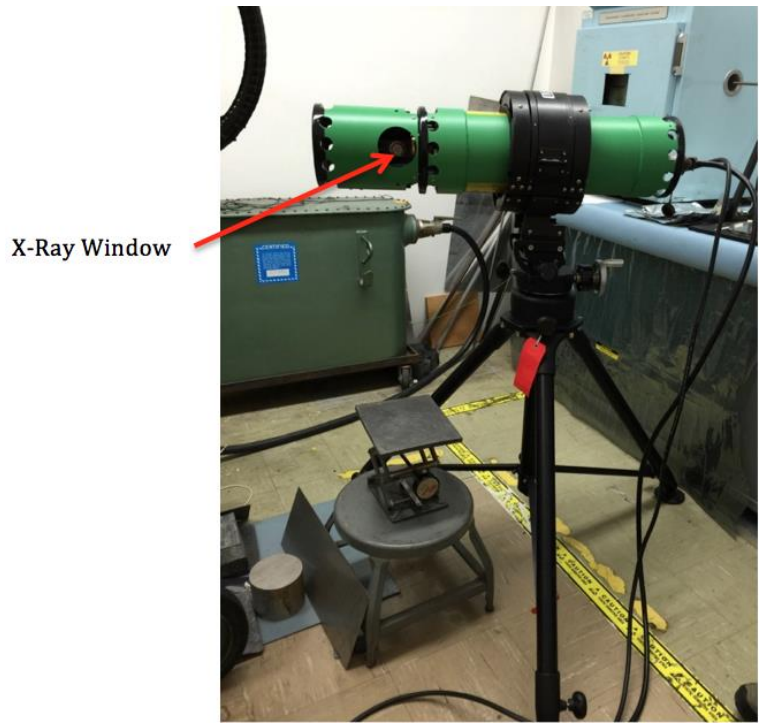

**Figure 5.** The LPX160 portable x-ray machine during setup. In this view, the body of the unit is rotated to show the beryllium window. During exposures the window faced downward toward the floor.

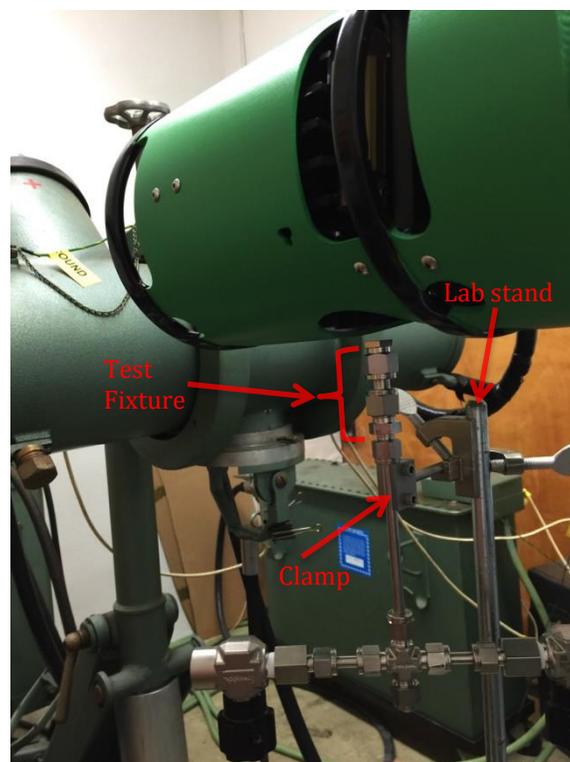

**Figure 6.** The test fixture is held by a clamp mounted to a vertical lab stand. In this view, the "cap" of the fixture is directly below the body of LPX160. Before irradiation was begun, the x-ray machine was lowered until the "cap" was nearly touching the beryllium window, as shown in the next figure.

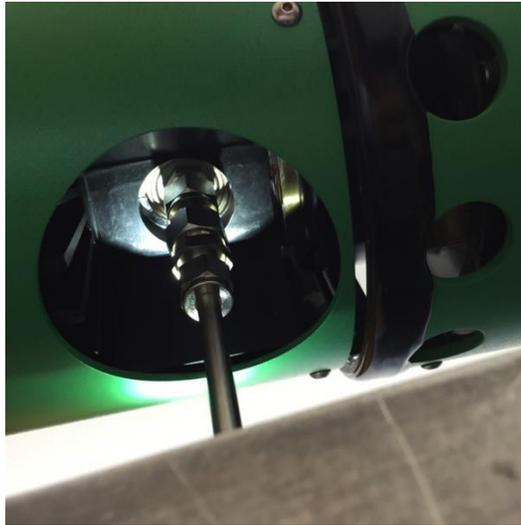

**Figure 7.** Looking up from below, the test fixture is shown positioned ready for exposure. The "cap" of the test fixture is approximately 1/16" from the beryllium window of the x-ray machine.

### 2.4.1. Beam Spot

One distinction between the LPX160 and the machine used in [1] is the size of the beam spot. Reference [1] describes the unit as a "microfocus" machine; whereas, the LPX160 is not microfocus. This difference is inconsequential to the results of the present research.

Microfocus is a term used to describe an x-ray machine with a small beam spot, typically on the order of 10's of micrometers. The advantage of microfocus is that it provides crisper x-ray images compared to non-microfocus. To understand why, consider Figure 8a, which depicts a rod with a crack that is being imaged with x-rays emanating from a single point. As a result of the single point source, each location along the crack appears in a single pixel in the imager, making for a clear image.

Contrast this with the situation shown in Figure 8b, where half of the x-rays emanate from one point source and the other half emanate from a second point source a small distance away. In this case, each location along the crack appears in two places on the imager. Undoubtedly, this is bad for image clarity, however, the crack in the rod still sees the same total flux of x-rays. Thus, in our case where x-ray flux is key, there is no reason to prefer microfocus to the 1.5mm beam spot provided by the LPX160.

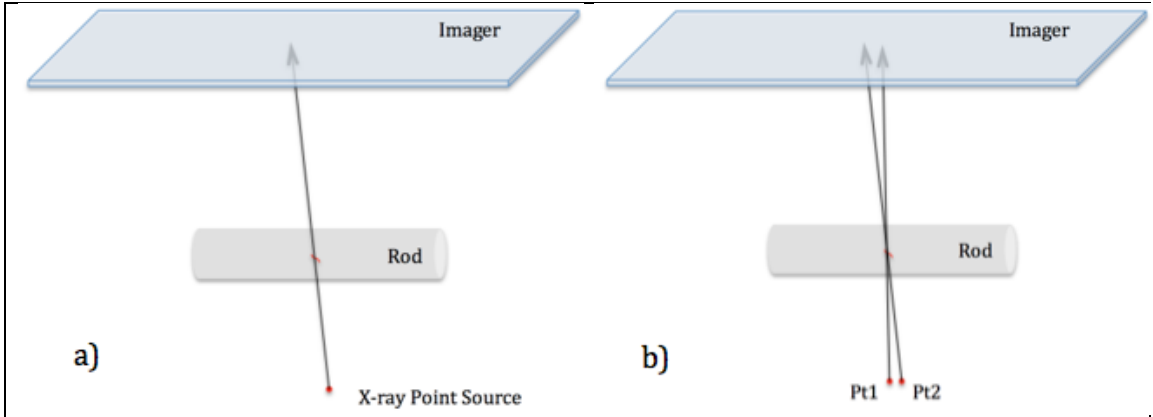

**Figure 8.** Depiction of a crack in a metal rod being imaged with a single x-ray point source (left) and two point sources (right).

### 2.4.2. Energy Spectrum

Figure 9 shows estimates of x-ray spectra for 160 keV and 200 keV accelerating potentials before and after the test fixture's stainless steel cap. The initial spectra were generated using SpekCalc [3][4][5]. In both cases, the spectrum is estimated at 3cm distance from a tungsten target with a 0.8mm beryllium window and 1mm air gap in between[**].

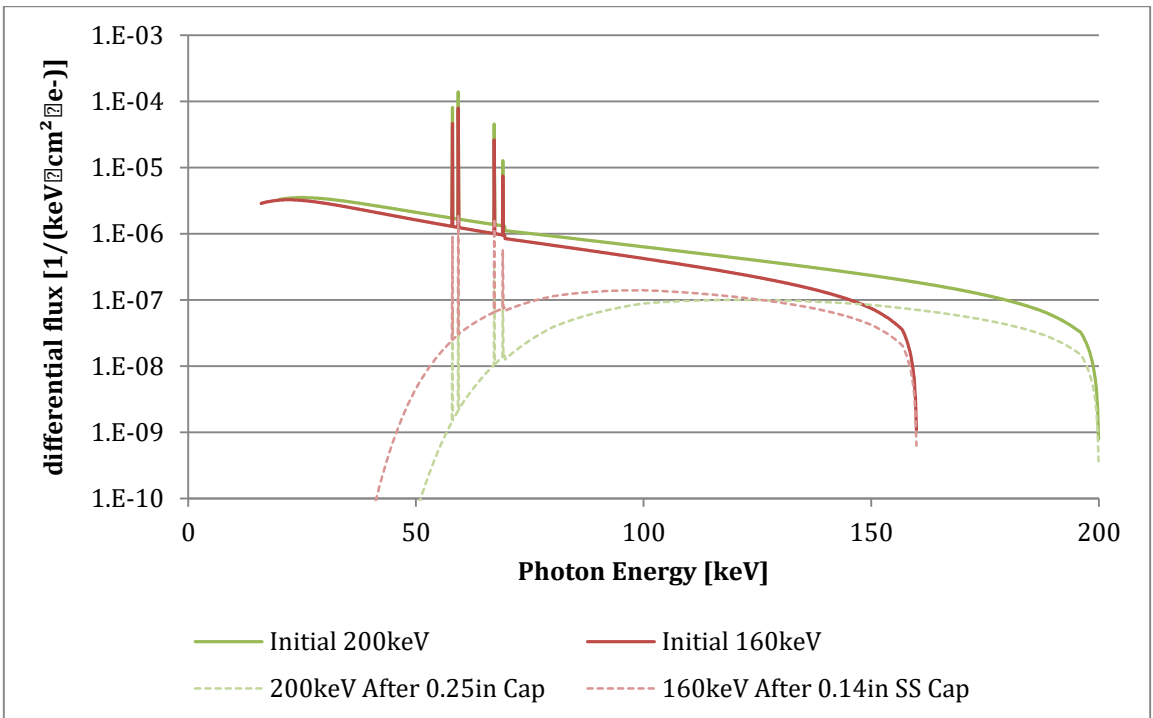

**Figure 9.** Energy spectra before and after stainless steel cap.

---

[**] The balance of the distance is vacuum to represent the inside of the x-ray tube.

An important message from Figure 9 is the value in having thinned the test fixture's cap. Before the shielding of the cap, the 200keV flux is greater than the 160keV spectrum at all energies. However, because of the thinned cap, the flux for our irradiations was greater than for samples from [1] for energies up to 125keV.

The effect of the stainless steel cap has been estimated using the standard attenuation relationship

$$I = I_0 e^{-\mu x}$$

where $I$ and $I_0$ are the initial and shielded x-ray beam intensities, $x$ is the thickness of the stainless steel, and $\mu$ is the attenuation coefficient. The attenuation coefficients as a function of energy were obtained from NIST's XCOM tool [6]. We used a stainless steel density of 8.03 g/cm$^3$, and the elemental composition given in Table 3. The resulting attenuation is shown in Figure 10.

**Table 3.** Assumed composition of stainless steel.

| Element | mass % |
|---|---|
| Ni | 0.12 |
| Cr | 0.17 |
| Mo | 0.025 |
| Si | 0.01 |
| Mn | 0.02 |
| Fe | 0.655 |

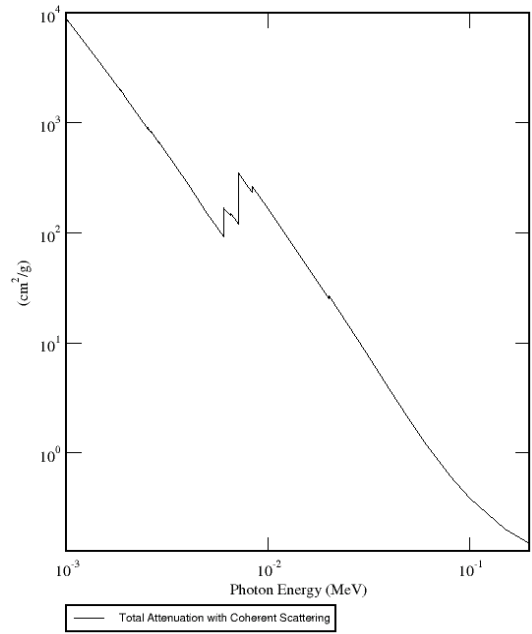

**Figure 10.** Stainless steel attenuation from 1keV to 200keV.

## 2.5. Detector

After x-ray exposure, the sample remained in the test fixture for the few minute walk to the Radiation Safety Office, which is home for JPL's low background alpha and beta detector, namely a Canberra Tennelec S5XLB. Once there, the fixture was opened and the material was placed in a planchet lined with double face tape, as discussed in Section 2.2.

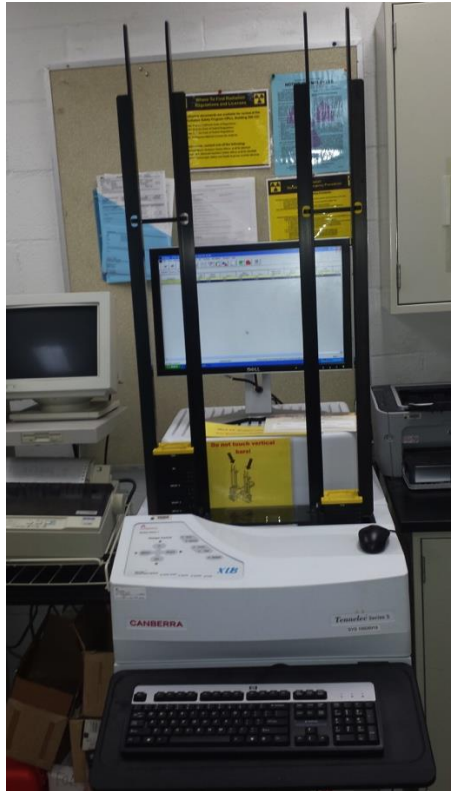

Figure 11. JPL's Canberra alpha/beta detector.

The Canberra is used frequently by JPL's Radiation Safety Office for analyzing wipe tests. The system was calibrated on July 9, 2015. The results of the calibration are summarized in Table 4.

Table 4. Summary of Canberra calibration results.

| Particle | Efficiency | Background |
|---|---|---|
| Alpha | 38.46% ± 0.24 | 0.03 ± 0.03 cpm |
| Beta | 47.74% ± 0.45 | 1.60 ± 0.23 cpm |

For the present work, 10 minute integration times were used for all measurements. A sample of the report that is generated for each run is shown in Figure 12.

![Figure 12 sample report table]

Figure 12. An example of a Canberra report. This is the report for Sample number 1 run immediately following x-ray exposure.

The Alpha CPM and Beta CPM are simply the number of counts detected divided by the 10 minutes integration time. The sample activity is then calculated using

$$Activity = \frac{CPM_{measured} - CPM_{background}}{Efficiency} \cdot 4.5E - 7$$

where efficiency, the ratio of CPM to DPM, was determine during the July 9 calibration with a sources[††] with known activity. The background (i.e. $CPM_{background}$) is also determined during calibration.

Between measurements, samples were covered with a second planchet and stored in a closed cabinet adjacent to the Canberra.

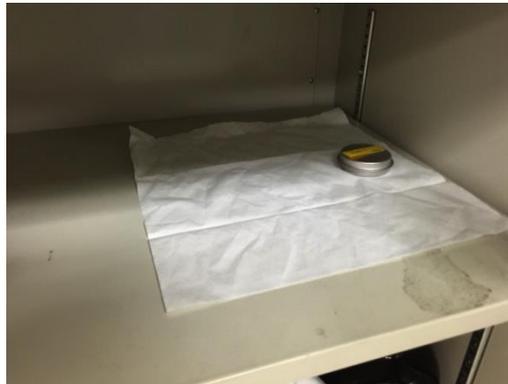

Figure 13. Storage of samples between measurements.

---

[††] An Am-241 source was used for alpha efficiency measurement and Sr-90 for beta.

## 3. Run Summary

Table 5 lists the conditions of each x-ray run. The odd numbered samples contained the deuterated target materials. The even numbered samples contained the control materials. For comparison, the x-ray exposure of Sample SL16 from [1] consisted of 1.0mA of current for 60 minutes at 200kV with 0.25" of stainless steel shielding.

Table 5. A summary of the conditions of the x-ray exposures. "SS"=Stainless Steel, "Al"=Aluminum

| ID | Material | Volts (kV) | Current (A) | Shielding | Duration (min) | Date (2015) | Completion |
|---|---|---|---|---|---|---|---|
| Sample #1 | Deuterium | 160 | 1.0 | 0.14" SS | 90 | 20-Aug | 10:40 AM |
| Sample #2 | Hydrogen | 160 | 1.0 | 0.14" SS | 90 | 20-Aug | 1:58 PM |
| Sample #3 | Deuterium | 160 | 1.0 | 0.14" SS | 90 | 24-Aug | 10:16 AM |
| Sample #4 | Hydrogen | 160 | 1.0 | 0.14" SS | 90 | 24-Aug | 1:15 PM |
| Sample #5 | Deuterium | 160 | 1.0 | 0.001" Al | 90 | 26-Aug | 10:17 AM |
| Sample #6 | Hydrogen | 160 | 1.0 | 0.001" Al | 90 | 26-Aug | 2:09 PM |
| Sample #7 | Deuterium | 160 | 2.0 | 0.001" Al | 90 | 27-Aug | 10:10 AM |
| Sample #8 | Hydrogen | 160 | 2.0 | 0.001" Al | 90 | 27-Aug | 2:01 PM |

## 4. Test Results

All alpha and beta activity measurements taken with the Canberra are provided in Appendix A4 (Table 7). Empty planchets ("blanks") were run frequently to alert the experimenter to any systematic shifts over time. Figure 14 shows the counts per minute for all blank measurements. There is no systematic shift, indicating no contamination or instrumentation shift.

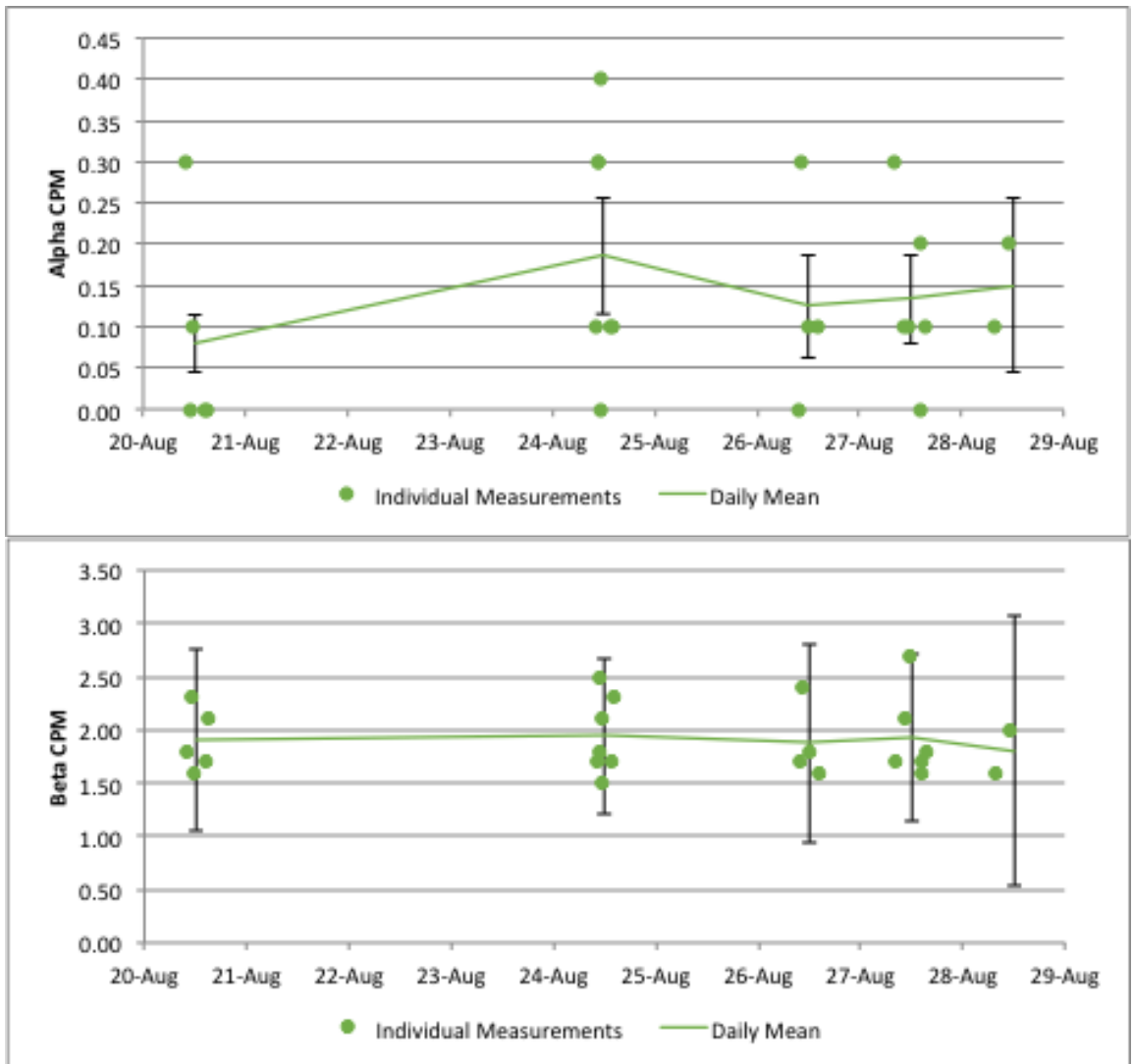

**Figure 14.** Counts per minute for blank planchets over time. Error bars represent error on the mean.

Each sample was measured soon after exposure[‡‡] and then multiple times over subsequent days. Figure 15 shows alpha and beta counts per minute measured over time. Table 1 on page 3 provides a different summary of the data. It focuses on the first measurement after each x-ray exposure ended. In both cases, the time sequence of Figure 15 and the initial measurements summarized in Table 1, there is no difference between the deuterated samples and control samples. Also, comparing Figure 15 and Figure 14 reveals that the average counts per minute for both alphas and betas immediately after irradiation and over time are at or below the counts for the blank planchets. Finally, as shown in Table 1, the measured activities (in decays per minute) are consistent with zero within uncertainties and are all below the

---

[‡‡] Table 1 contains the actual lag time between exposure and measurement.

Minimum Detectable Amount (MDA). All of these observations are consistent with no activation of the target materials.

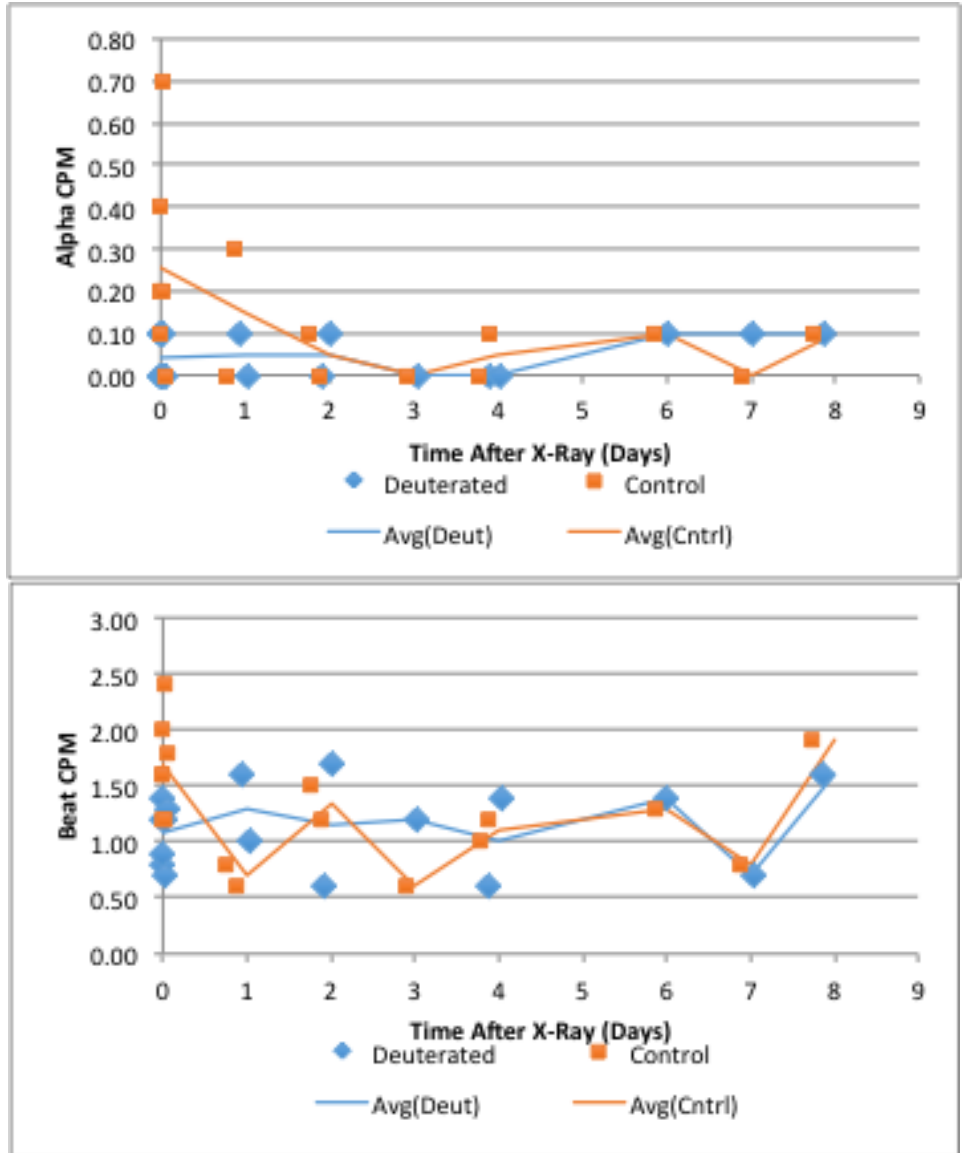

**Figure 15.** Counts per minute for deuterated and control samples over time.

## 5. Severity of Run Conditions

In this section, we calculate one metric that may be of value in comparing the severity of our run conditions to those in [1]. Namely, for each run condition, we estimate the number of photon-electron interactions in the target material as a function of energy.

Certainly, more thorough analyses are possible, particularly if one includes a hypothesis for the physics that leads to the alpha and beta decays observed in [1].

Regardless of the theory conjectured, however, the physical chain of events, presumably, begins with the basic x-ray energy deposition processes: Compton scattering and photoelectric absorption. Therefore, we present the number of interactions as a way of gauging run severity.

The calculation of the number of photon-electron interactions uses the same beam attenuation approach used as in Section 2.4.2 except this time the photons are interacting in the PE and titanium based sample materials. Again we use XCOM to generate shielding coefficients. For the 0.5g $TiD_{1.97}$ / 0.5g DPE mixture, we calculate an average density of 2.48 g/cm³ and use the elemental compositions shown in Table 6 to yield the attenuation coefficient profile shown in Figure 16. Similarly, for the hydrogen-based control sample, a density of 2.34 g/cm3 and composition shown in Table 6 were used to calculate attenuation coefficients shown in Figure 16.

Table 6. Composition of TiD/DPE (left) and TiH/PE (right) used in calculating attenuation coefficients.

| Element | mass % |
|---------|--------|
| D       | 16.35  |
| C       | 37.5   |
| Ti      | 46.144 |

| Element | mass % |
|---------|--------|
| H       | 9.14   |
| C       | 42.8   |
| Ti      | 47.99  |

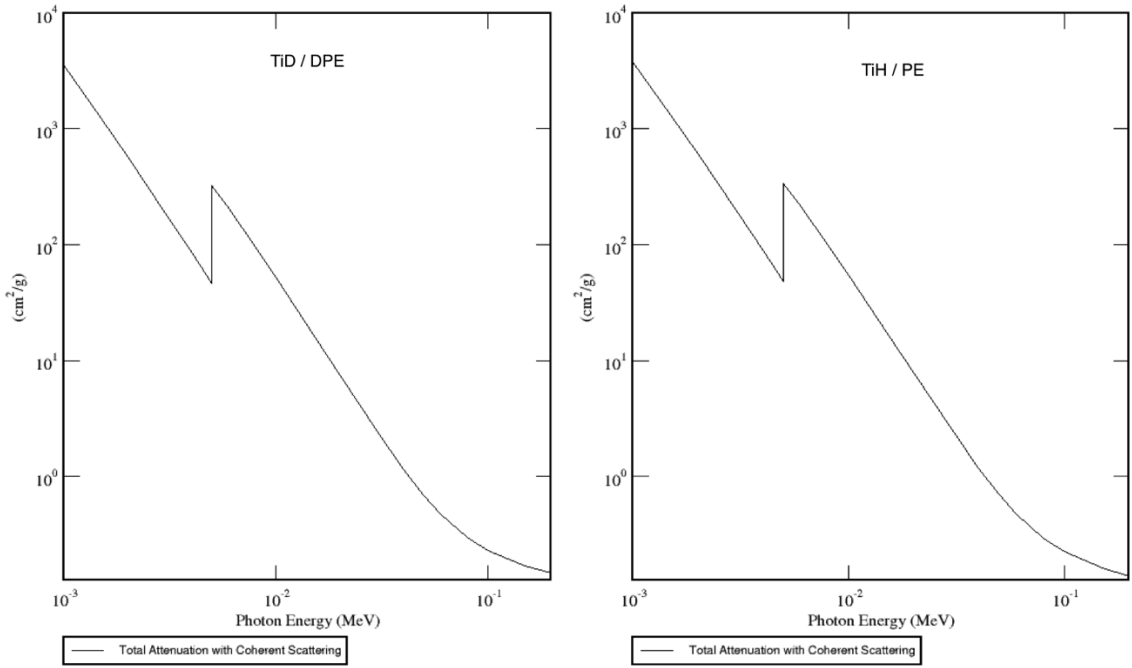

Figure 16. Attenuation of TiD / DPE (left) and TiH / PE (right) mixtures.

To first order, the number of photon-electron interactions in the sample material is equal to the number of primary-beam photons entering the sample minus the number exiting. In each case, the number is equal to the beam intensity multiplied by the area of sample (as viewed down the long axis of the text fixture) and the duration of the exposure.

$$N = (I_{enter} - I_{exit}) \cdot A \cdot t$$

The numbers entering and exiting are calculated using the shielding formula discussed in Section 2.4.2.

$$N = [I_0 e^{-\mu_{shielding} x_{shielding}} (1 - e^{-\mu_{sample} x_{sample}})] \cdot A \cdot t$$

The shielding thickness, material, and duration for each run is given in Table 5. The area of the sample is the cross-sectional area of the sample holder in the test fixture. Figure 17 shows the resulting number of photons interacting in the material. The plot shows integral number above a given photon energy. In terms of total number of interactions all of the runs from the present work were more severe than the SL16 run from [1]. Nonetheless, we do miss out on the highest energy photons, so none of our runs is as severe as those in [1] for processes activated by photons >144keV.

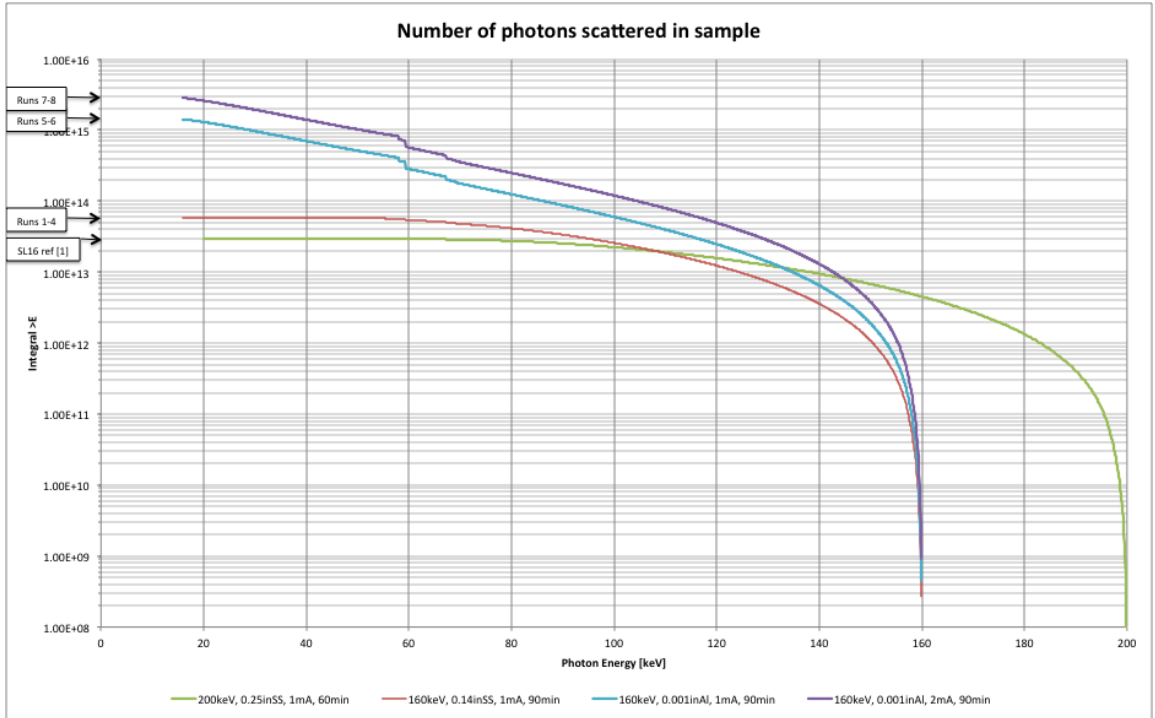

**Figure 17.** Number of photons scattered in sample.

## 6. Conclusion

We find no evidence of x-ray induced nuclear activation of deuterated materials with 160keV tungsten-target x-rays. Our run conditions probe processes initiated by scattering of photons up to 144keV better than the run conditions for sample SL16 in [1]. Thus, either the activity observed by the previous researchers has an energy above 160keV, required a greater fluence than we created between 144 and 160 keV, or an alternate explanation exits for the activity observed in [1].

The alphas reported in [1] are quite perplexing and difficult to explain given the target materials and photon energies involved. None of the isotopes in the target materials are alpha emitters, nor are there alpha emitters nearby on the chart of nuclides. We conclude that the most likely explanation for the observed alpha activity is contamination. Furthermore, if alpha emitters could be introduced through contamination, then it seems that beta emitters might also.

## 7. Alternate Hypothesis

Given our null result, we propose, as an alternate hypothesis, that the materials in the previous work were contaminated with daughters of radon decay. Radon gas ($^{222}$Rn) occurs naturally as part of the uranium decay chain and is found in many parts of the United States. See Figure 18.

As discussed in [8] and references therein, radon decay daughters can accumulate on electrostatically charged dielectrics. If the polyethylene materials became electrostatically charged during preparation and handling, they may have attracted radon decay daughters subsequently. The plausibility of this theory is enhanced by the fact that the work in [1] was conducted partially in a basement laboratory because radon gas tends to concentrate in basements.

As one can see in Figure 19, the daughters of radon decay contain both alpha and beta emitters. It is interesting to note that the first 5 steps of the chain, from $^{218}$Po to $^{210}$Pb, contain alphas and has a total half-life of 50 minutes. This is followed by $^{210}$Pb decaying by beta decay, with a 22year half-life. Thus, the 214Po →210Pb alpha would be around for the first few hours but would disappear after that. The final alpha in the chain $^{210}$Po → $^{206}$Pb wouldn't show up for years. This timeline seems consistent with the observation in [1] that alphas were present for the first measurement at about ½ hour after exposure, but had disappeared during subsequent measurements.

Both the PE and DPE would be capable of introducing the radon daughter contamination. Thus, the fact that activity was never observed from the control materials in [1], represents a potential flaw with our hypothesis. The difference might be explained by different processing methods or timing of the control materials as compared to the target materials.

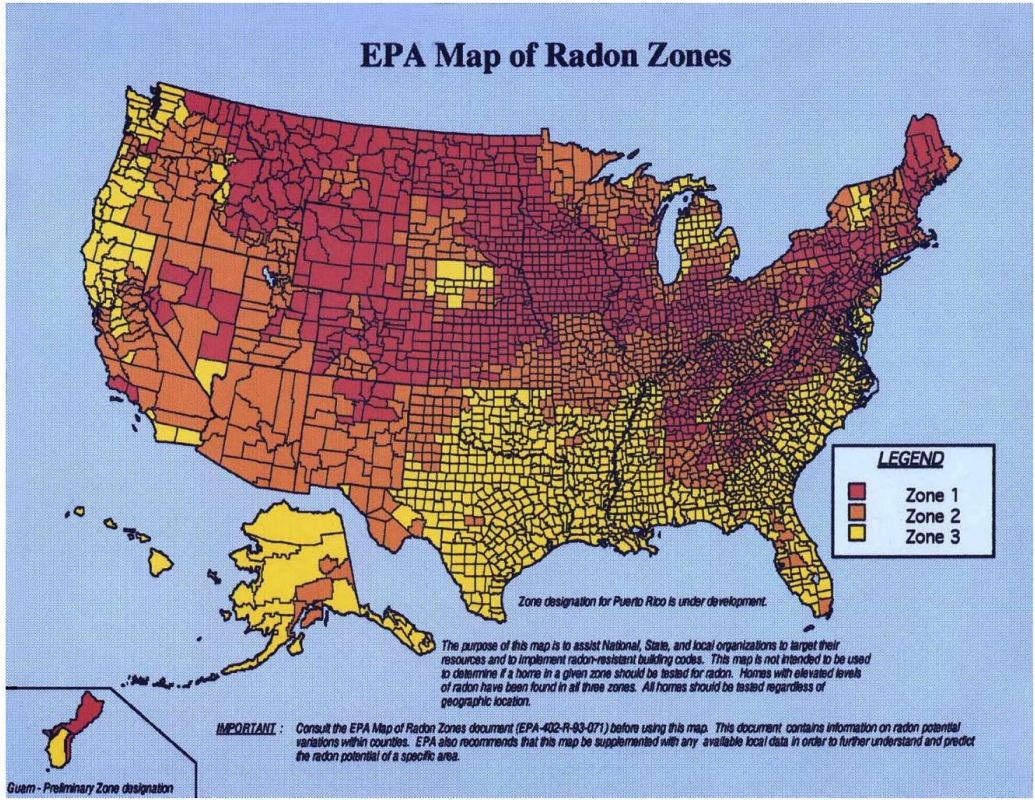

**Figure 18.** Radon Zones across the US [7]. Zones 1, 2, 3 → Highest, Moderate, Low Potential.

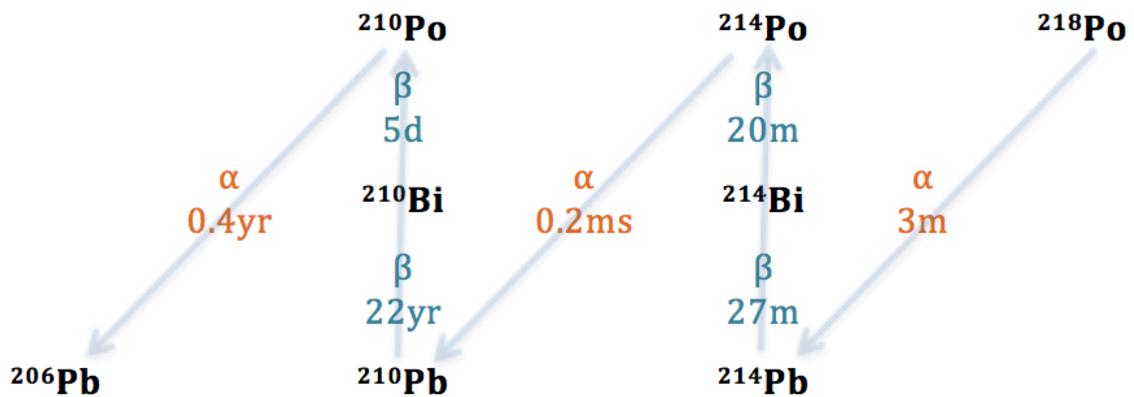

**Figure 19.** Daughters of radon gas decay. Radon gas ($^{222}$Rn) decays directly to $^{218}$Po, which is shown in the upper right of the figure. It, in turn, decays via alpha decay with a 3-minute half-life to 214Pb, and so on.

## 7.1. Electrostatic Charging Investigation

The radon daughter hypothesis led us to question whether the sample materials could easily become charged and retain charge. Consequently, we used a Simco FMX-003 Electrostatic Field Meter, to measure the relative charge state of samples of the PE and DPE under various circumstances. Specifically, we did the following:
1. Measured electrostatic charge of PE and DPE, untouched from the manufacturer.
2. Investigated whether chopping material with a razor changes the charging level of the material
3. Explored means of discharging the charged material with aluminum foil and an air ionizer.

We found that both PE and DPE were relatively easily charged, with indications that the PE was somewhat more susceptible. The charge state was still present after 36 hours, but was gone a few days after that. Attempts to remove the charge from some samples through grounding with grounded sheet of aluminum foil had no effect. The charge on the material could, however, be neutralized by using an ion blower.

This was just a preliminary investigation into how easily these materials could become charged. We found that the materials can become charged and hold a charge for a period of days. While this does not prove the radon daughter hypothesis, it does demonstrate that there is a physical mechanism that could have contributed to contamination of the sample.

## 8. Recommendations

Our null result raises doubts about the claim of nuclear activation by x-rays reported in [1]. If their work is to be pursued further, we suggest the following improvements to the experimental technique.
- Samples should be stabilized (as we did with double faced tape) to enable repeatable measurements.
- Measure activity over time and attempt to identify ½ lives of constituent decays.
- Use gamma spectroscopy to look for gamma's associated with beta decay.
- Measure electrostatic field of sample materials. Use an ion blower to neutralize charge.
- Measure Radon concentration in the experimental area.

We also note the following ways in which the analyses in the present report could be supplemented.
1. Use energy absorption coefficients to estimate total energy deposited.
2. Monte Carlo modeling with MCNPX, GEANT4, or similar could be used to calculate total number of electrons generated, the energy spectrum of all electrons generated, effects of backscattering, and to improve the accuracy of the results through improved geometry.

Appendix

# Appendix A1: Manufacturer's Information on Polyethylene Materials

**Sample Name**: Polyethylene
(obtained from the hydrogenation of Poly butadiene rich in 1,4 microstructure)

**Sample #**: P1572-E

**Structure**:

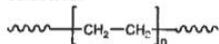

**Composition**:

| Mn x 10³ | PDI |
|---|---|
| 114.0 | 1.10 |

**Synthesis Procedure**:
Polyethylene is made from the hydrogenation of 1,4-polybutadiene. 1,4-polybutadiene is synthesized by by living anionic polymerization of butadiene in non-polar solvent.

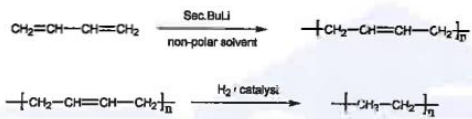

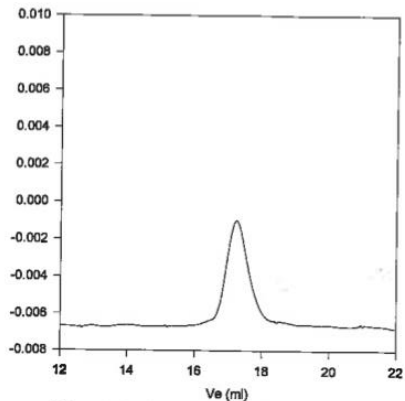

SEC of the Polymer: Precursor
P1445-Bd (Precursor for P1572-E)

Size exclusion chromatography of polybutadiene:
$M_n$=110,000, $M_w$=121,000 PI=1.10
PE after Hydrogenation : Mn 114,000

**Characterization**:
The molecular weight and polydispersity index (PDI) are obtained by size exclusion chromatography (SEC) in THF. The SEC instrument calibrated with poly butadiene standards. SEC analysis was performed on a Varian liquid chromatograph equipped with refractive and UV light scattering detectors. Three SEC columns from Supelco (G6000-4000-2000 HXL) were used with triple detectors from Viscotek Co.

The hydrogenation of polybutadiene is confirmed by FT-IR with disappearance of the alkene double bond.

**Solubility**:
Polyethylene is soluble in hot toluene and hot xylene. The polymer is insoluble in hexane, methanol and ethers.

124 Avro Street, Dorval (Montréal) PQ. H9P 2X8 CANADA
Toll Free: 1-866-422-9842 Voice: 514-421-5517 Fax: 514-421-5518
e-mail: contact@polymersource.com web-site: http://www.polymersource.com

**Sample Name:** Polyethylene (obtained from the hydrogenation of Poly butadiene rich in 1,4 microstructure)

**Sample #:** P2250-E

**Structure:**

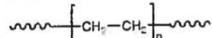

**Composition:**

| Mn × 10³ | PDI |
|---|---|
| 103.2 | 1.08 |

| $T_m$ (°C): 98 | $T_c$ (°C): 77 | $T_g$ (°C): - |
|---|---|---|

**Synthesis Procedure:**
Polyethylene is made from the hydrogenation of 1,4-polybutadiene. 1,4-polybutadiene is synthesized by by living anionic polymerization of butadiene in non-polar solvent.

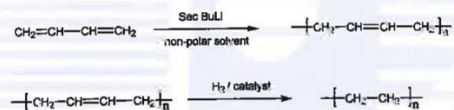

**Characterization:**

The molecular weight and polydispersity index (PDI) are obtained by size exclusion chromatography. The hydrogenation of polybutadiene is confirmed by FT-IR with disappearance of the alkene double bond.
Thermal analysis of the samples was carried out on a TA Q100 differential scanning calorimeter at a heating rate of 10°C/min.
The melting temperature ($T_m$) was taken as the maximum of the endothermic peak where as the crystallization temperature ($T_c$) was considered as the minimum of the exothermic peak.

**Solubility:**
Polyethylene is soluble in hot toluene and hot xylene. The polymer is insoluble in hexane, methanol and ethers.

**SEC of the Polymer: Precursor**

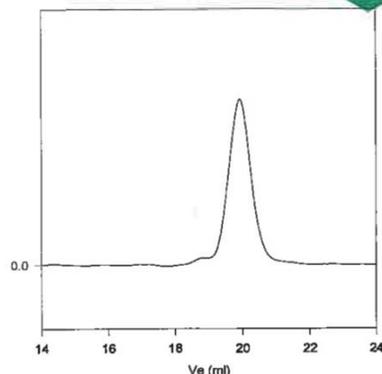

Size exclusion chromatography of polybutadiene with respect to polybutadiene standards (precursor for P1990-E):
$M_n$=99500, $M_w$=107500, $M_w/M_n$=1.08
Molecular weight of Polyethylene Mn 103200 Mw/Mn:1.08

**Melting curve for the sample:**

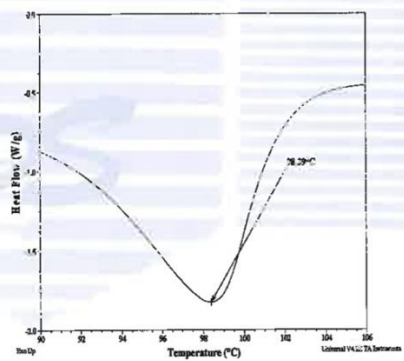

**Crystallization curve for the sample:**

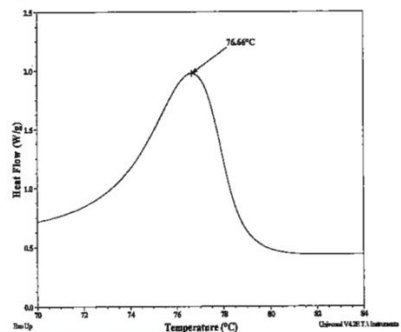

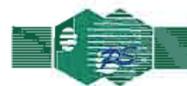

# CERTIFICATE OF ANALYSIS

**Sample Name:** Deuterated Polyethylene

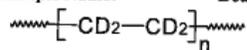

**Chemical Purity Specification:** $\geq 98\%$

**Labeled CAS Number:** 25549-98-8

**Unlabeled CAS Number:** 9002-88-4

**Molecular Weight:** $M_w$ 361,000; $M_w/M_n$ 1.7

**Chemical Formula:** $(CD_2CD_2)n$

**Storage:** Store at room temperature.
**Stability:** Stable at room temperature

**Sample Lot No.:** P9549-dPE

**Appearance:** White chunks and powder but when compressed molded it turns to light yellow or ivory color mold

**Method of analysis:** Size exclusion chromatography

Appendix A2: JPL's Confirmation on Quality of the DPE

*JPL* ANALYTICAL CHEMISTRY LABORATORY                                    ACL-A050
*Analytical Chemistry and Materials Development Group 3531*
*Propulsion and Materials Section 3530*

---

**To:**        Emma Bradford                                              8/26/2015

From:        Mark S. Anderson

Subject:    Analysis of deuterated polyethylene for undeuterated polyethylene impurity

**Purpose**

A sample of deuterated polyethylene was submitted to determine if there was undeuterated polyethylene in the sample. There was a need for >98% deuterated polyethylene purity.

**Results and Discussion**

The amount of polyethylene in the deuterated polyethylene is less than 1%. The infrared and Raman spectra both show minor amounts of C-H relative to the D-H peaks of the deuterated polyethylene. There are small amounts of C-H from trace amounts of hydrocarbon impurities. This analysis did not measure non-polymeric fillers.

**Figure**: Infrared spectra of deuterated and a reference polyethylene (a, b) and corresponding Raman spectra (c,d).

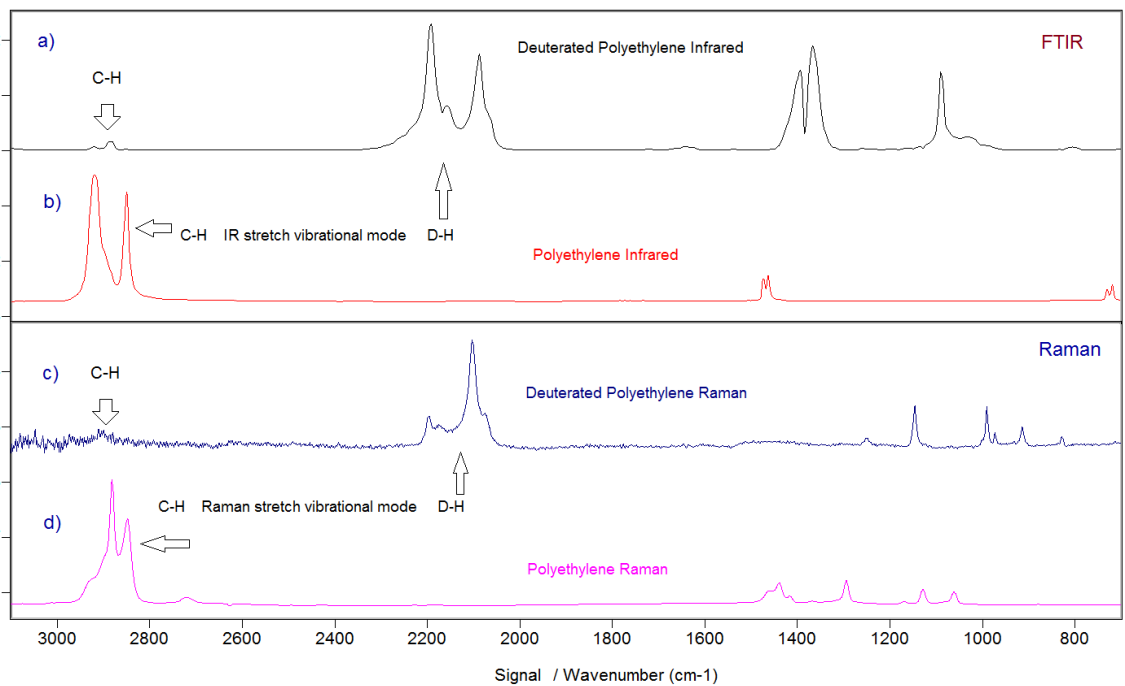

**Method**

The amount of polyethylene (undeuterated) in the deuterated polyethylene sample was measured using vibrational spectroscopy (infrared and Raman). This was done by measuring the ratio of C-H peaks of undeuterated polyethylene to the deuterated polyethylene C-D stretch peaks. The frequency of the C-D peaks are sufficiently shifted to provide discrimination and an estimation of the amount of polyethylene in the deuterated sample. The extinction coefficients and/or Raman cross sections are assumed to be approximately the same for C-H and C-D.

The material was analyzed using diffuse reflectance, Fourier Transform Infrared Spectroscopy (DRIFT-FTIR). FTIR provides chemical functional group information for quantitative analysis and qualitative identification of materials.

The material was analyzed using Raman Micro-Probe. Raman spectroscopy provides chemical functional group information for quantitative analysis and qualitative identification of materials. The Raman system is a Bruker Senterria microprobe with a ~5 micron spot size using 785nm excitation (2mW) and 5 second integration time with 10 coadditions.

Appendix A3: Excerpt from X-Ray Machine's Datasheet

The LPX160 datasheet was removed from this appendix because it is copyrighted. It is available from the manufacturer's website:
http://www.spellmanhv.com/en/Products/Product-NDT/LPX160.aspx

# Appendix A4: Alpha and Beta Activity Measurements from Canberra

Table 7. Alpha and beta activity measurements from Canberra. Alpha and beta background from July 9 calibration: 0.03±0.03 cpm; 1.60 ± 0.23 cpm.

| Sample | Date (2015) | Report Time-stamp | Count Time (min) | Alpha CPM | Alpha DPM | Alpha DPM Unc. | Alpha Activity (µCi) | Alpha MDA | Beta CPM | Beta DPM | Beta DPM Unc | Beta Activity (µCi) | Beta MDA |
|---|---|---|---|---|---|---|---|---|---|---|---|---|---|
| Blank | 20-Aug | 10:08 AM | 10 | 0.30 | 0.69 | 0.46 | 0.00 | 1.27 | 1.80 | 0.42 | 1.01 | 0.00 | 3.75 |
| Sample #1 | 20-Aug | 10:57 AM | 10 | 0.00 | -0.09 | 0.09 | 0.00 | 1.27 | 1.40 | -0.42 | 0.92 | 0.00 | 3.75 |
| Blank | 20-Aug | 10:57 AM | 10 | 0.00 | -0.09 | 0.09 | 0.00 | 1.27 | 2.30 | 1.47 | 1.12 | 0.00 | 3.75 |
| Sample #1 | 20-Aug | 11:34 AM | 10 | 0.10 | 0.17 | 0.27 | 0.00 | 1.27 | 0.70 | -1.89 | 0.74 | 0.00 | 3.75 |
| Blank | 20-Aug | 11:47 AM | 10 | 0.10 | 0.17 | 0.27 | 0.00 | 1.27 | 1.60 | 0.00 | 0.97 | 0.00 | 3.75 |
| Sample #2 | 20-Aug | 2:13 PM | 10 | 0.20 | 0.43 | 0.38 | 0.00 | 1.27 | 1.60 | 2.09 | 1.17 | 0.00 | 3.75 |
| Blank | 20-Aug | 2:27 PM | 10 | 0.00 | -0.09 | 0.09 | 0.00 | 1.27 | 1.70 | 0.21 | 0.99 | 0.00 | 3.75 |
| Sample #2 | 20-Aug | 2:39 PM | 10 | 0.70 | 1.73 | 0.69 | 0.00 | 1.27 | 2.40 | 1.68 | 1.13 | 0.00 | 3.75 |
| Blank | 20-Aug | 2:52 PM | 10 | 0.00 | -0.09 | 0.09 | 0.00 | 1.27 | 2.10 | 1.05 | 1.07 | 0.00 | 3.75 |
| Blank | 24-Aug | 10:15 AM | 10 | 0.10 | 0.17 | 0.27 | 0.00 | 1.27 | 1.70 | 0.21 | 0.99 | 0.00 | 3.75 |
| Sample #3 | 24-Aug | 10:29 AM | 10 | 0.10 | 0.17 | 0.27 | 0.00 | 1.27 | 0.80 | -1.68 | 0.77 | 0.00 | 3.75 |
| Blank | 24-Aug | 10:29 AM | 10 | 0.30 | 0.69 | 0.46 | 0.00 | 1.27 | 1.80 | 0.42 | 1.01 | 0.00 | 3.75 |
| Sample #3 | 24-Aug | 10:56 AM | 10 | 0.00 | -0.09 | 0.09 | 0.00 | 1.27 | 1.20 | -0.84 | 0.87 | 0.00 | 3.75 |
| Blank | 24-Aug | 10:56 AM | 10 | 0.30 | 0.69 | 0.46 | 0.00 | 1.27 | 2.50 | 1.89 | 1.15 | 0.00 | 3.75 |
| Sample #1 | 24-Aug | 11:23 AM | 10 | 0.00 | -0.09 | 0.09 | 0.00 | 1.27 | 1.40 | -0.42 | 0.92 | 0.00 | 3.75 |
| Blank | 24-Aug | 11:23 AM | 10 | 0.40 | 0.95 | 0.53 | 0.00 | 1.27 | 2.10 | 1.05 | 1.07 | 0.00 | 3.75 |
| Sample #2 | 24-Aug | 11:23 AM | 10 | 0.10 | 0.17 | 0.27 | 0.00 | 1.27 | 1.20 | -0.84 | 0.87 | 0.00 | 3.75 |
| Blank | 24-Aug | 11:23 AM | 10 | 0.00 | -0.09 | 0.09 | 0.00 | 1.27 | 1.50 | -0.21 | 0.94 | 0.00 | 3.75 |
| Sample #4 | 24-Aug | 1:33 PM | 10 | 0.10 | 0.17 | 27.00 | 0.00 | 1.27 | 1.60 | 0.00 | 0.97 | 0.00 | 3.75 |
| Blank | 24-Aug | 1:34 PM | 10 | 0.10 | 0.17 | 0.27 | 0.00 | 1.27 | 1.70 | 0.21 | 0.99 | 0.00 | 3.75 |
| Sample #4 | 24-Aug | 1:56 PM | 10 | 0.20 | 0.43 | 0.36 | 0.00 | 1.27 | 1.20 | -0.84 | 0.87 | 0.00 | 3.75 |
| Blank | 24-Aug | 1:57 PM | 10 | 0.10 | 0.17 | 0.27 | 0.00 | 1.27 | 2.30 | 1.47 | 1.12 | 0.00 | 3.75 |
| Blank | 26-Aug | 9:44 AM | 10 | 0.00 | -0.09 | 0.09 | 0.00 | 1.27 | 1.70 | 0.21 | 0.99 | 0.00 | 3.75 |

| Sample | Date (2015) | Report Time-stamp | Count Time (min) | Alpha CPM | Alpha DPM | Alpha DPM Unc. | Alpha Activity (µCi) | Alpha MDA | Beta CPM | Beta DPM | Beta DPM Unc | Beta Activity (µCi) | Beta MDA |
|---|---|---|---|---|---|---|---|---|---|---|---|---|---|
| Sample #5 | 26-Aug | 10:31 AM | 10 | 0.00 | -0.09 | 0.09 | 0.00 | 1.27 | 0.90 | -1.47 | 79.00 | 0.00 | 3.75 |
| Blank | 26-Aug | 10:31 AM | 10 | 0.30 | 0.69 | 0.46 | 0.00 | 1.27 | 2.40 | 1.68 | 1.13 | 0.00 | 3.75 |
| Sample #1 | 26-Aug | 10:44 AM | 10 | 0.10 | 0.17 | 0.27 | 0.00 | 1.27 | 1.40 | -0.42 | 0.92 | 0.00 | 3.75 |
| Sample #2 | 26-Aug | 10:44 AM | 10 | 0.10 | 0.17 | 0.27 | 0.00 | 1.27 | 1.30 | -0.63 | 0.90 | 0.00 | 3.75 |
| Sample #3 | 26-Aug | 10:44 AM | 10 | 0.10 | 0.17 | 0.27 | 0.00 | 1.27 | 1.70 | 0.21 | 0.99 | 0.00 | 3.75 |
| Sample #4 | 26-Aug | 10:44 AM | 10 | 0.00 | -0.09 | 0.09 | 0.00 | 1.27 | 1.20 | -0.84 | 0.87 | 0.00 | 3.75 |
| Blank | 26-Aug | 11:50 AM | 10 | 0.10 | 0.17 | 0.27 | 0.00 | 1.27 | 1.80 | 0.42 | 1.01 | 0.00 | 3.75 |
| Sample #6 | 26-Aug | 2:22 PM | 10 | 0.20 | 0.43 | 0.38 | 0.00 | 1.27 | 1.20 | -0.84 | 0.87 | 0.00 | 3.75 |
| Blank | 26-Aug | 2:22 PM | 10 | 0.10 | 0.17 | 0.27 | 0.00 | 1.27 | 1.60 | 0.00 | 0.97 | 0.00 | 3.75 |
| Blank | 27-Aug | 8:04 AM | 10 | 0.30 | 0.69 | 0.46 | 0.00 | 1.27 | 1.70 | 0.21 | 0.99 | 0.00 | 3.75 |
| Sample #7 | 27-Aug | 10:21 AM | 10 | 0.10 | 0.17 | 0.27 | 0.00 | 1.27 | 1.20 | -0.84 | 0.87 | 0.00 | 3.75 |
| Blank | 27-Aug | 10:22 AM | 10 | 0.10 | 0.17 | 0.27 | 0.00 | 1.27 | 2.10 | 1.05 | 1.07 | 0.00 | 3.75 |
| Sample #7 | 27-Aug | 11:24 AM | 10 | 0.00 | -0.09 | 0.09 | 0.00 | 1.27 | 1.30 | -0.63 | 0.90 | 0.00 | 3.75 |
| Blank | 27-Aug | 11:25 AM | 10 | 0.10 | 0.17 | 0.27 | 0.00 | 1.27 | 2.70 | 2.30 | 1.19 | 0.00 | 3.75 |
| Sample #1 | 27-Aug | 11:25 AM | 10 | 0.10 | 0.17 | 0.27 | 0.00 | 1.27 | 0.70 | -1.89 | 0.74 | 0.00 | 3.75 |
| Sample #2 | 27-Aug | 11:25 AM | 10 | 0.00 | -0.09 | 0.09 | 0.00 | 1.27 | 0.80 | -1.68 | 0.77 | 0.00 | 3.75 |
| Sample #3 | 27-Aug | 11:25 AM | 10 | 0.00 | -0.09 | 0.09 | 0.00 | 1.27 | 1.20 | -0.84 | 0.85 | 0.00 | 3.75 |
| Sample #4 | 27-Aug | 11:25 AM | 10 | 0.00 | -0.09 | 0.09 | 0.00 | 1.27 | 0.60 | -2.09 | 0.71 | 0.00 | 3.75 |
| Sample #5 | 27-Aug | 11:25 AM | 10 | 0.00 | -0.09 | 0.09 | 0.00 | 1.27 | 1.00 | -1.26 | 0.82 | 0.00 | 3.75 |
| Sample #6 | 27-Aug | 11:25 AM | 10 | 0.30 | 0.69 | 0.46 | 0.00 | 1.27 | 0.60 | -2.09 | 0.71 | 0.00 | 3.75 |
| Blank | 27-Aug | 2:04 PM | 10 | 0.20 | 0.43 | 0.38 | 0.00 | 1.27 | 1.60 | 0.00 | 0.97 | 0.00 | 3.75 |
| Sample #8 | 27-Aug | 2:22 PM | 10 | 0.40 | 0.95 | 0.53 | 0.00 | 1.27 | 2.00 | 0.84 | 1.05 | 0.00 | 3.75 |
| Blank | 27-Aug | 2:22 PM | 10 | 0.00 | -0.09 | 0.09 | 0.00 | 1.27 | 1.70 | 0.21 | 0.99 | 0.00 | 3.75 |
| Sample #8 | 27-Aug | 3:18 PM | 10 | 0.00 | -0.09 | 0.09 | 0.00 | 1.27 | 1.80 | 0.42 | 1.01 | 0.00 | 3.75 |
| Blank | 27-Aug | 3:18 PM | 10 | 0.10 | 0.17 | 0.27 | 0.00 | 1.27 | 1.80 | 0.42 | 1.01 | 0.00 | 3.75 |

| Sample | Date (2015) | Report Time-stamp | Count Time (min) | Alpha CPM | Alpha DPM | Alpha DPM Unc. | Alpha Activity (µCi) | Alpha MDA | Beta CPM | Beta DPM | Beta DPM Unc | Beta Activity (µCi) | Beta MDA |
|---|---|---|---|---|---|---|---|---|---|---|---|---|---|
| Blank | 28-Aug | 7:28 AM | 10 | 0.10 | 0.17 | 0.27 | 0.00 | 1.27 | 1.60 | 0.00 | 0.97 | 0.00 | 3.75 |
| Sample #1 | 28-Aug | 7:32 AM | 10 | 0.10 | 0.17 | 0.27 | 0.00 | 1.27 | 1.60 | 0.00 | 0.97 | 0.00 | 3.75 |
| Sample #2 | 28-Aug | 7:32 AM | 10 | 0.10 | 0.17 | 0.27 | 0.00 | 1.27 | 1.90 | 0.63 | 1.03 | 0.00 | 3.75 |
| Sample #3 | 28-Aug | 7:32 AM | 10 | 0.00 | -0.09 | 0.09 | 0.00 | 1.27 | 0.60 | -2.09 | 0.71 | 0.00 | 3.75 |
| Sample #4 | 28-Aug | 7:32 AM | 10 | 0.00 | -0.09 | 0.09 | 0.00 | 1.27 | 1.00 | -1.26 | 0.82 | 0.00 | 3.75 |
| Sample #5 | 28-Aug | 7:32 AM | 10 | 0.00 | -0.09 | 0.09 | 0.00 | 1.27 | 0.60 | -2.09 | 0.71 | 0.00 | 3.75 |
| Sample #6 | 28-Aug | 7:32 AM | 10 | 0.10 | 0.17 | 0.27 | 0.00 | 1.27 | 1.50 | -0.21 | 0.94 | 0.00 | 3.75 |
| Sample #7 | 28-Aug | 7:32 AM | 10 | 0.10 | 0.17 | 0.27 | 0.00 | 1.27 | 1.60 | 0.00 | 0.97 | 0.00 | 3.75 |
| Sample #8 | 28-Aug | 7:32 AM | 10 | 0.00 | -0.09 | 0.09 | 0.00 | 1.27 | 0.80 | -1.68 | 0.77 | 0.00 | 3.75 |
| Blank | 28-Aug | 11:00 AM | 10 | 0.20 | 0.43 | 0.38 | 0.00 | 1.27 | 2.00 | 0.84 | 1.05 | 0.00 | 3.75 |

# Appendix A5: Extension of results to 200kV

## 10. Test Setup
## 10.1. Material Sample
All samples for the additional measurements were made from unused material remaining from August 2015. It was prepared in the same manner as described in Section 2.2.

## 10.2. X-Ray Equipment
The x-ray irradiations were performed at North Star Imaging in Irvine, CA. The x-ray tube in North Star's equipment is a FeinFocus FXE 225. The tube used in [1] was an X-Ray WorX XWT-225-SE. Both tubes use tungsten targets and are very similar in design and construction. For instance, the distance from the tungsten target to the exit window for the FeinFocus is 6.75mm while [1] indicates that target to window distance is "~6mm" for their X-Ray WorX tube. Under our test conditions, the microfocus beam spot for the FeinFocus is 200um. All 4 of our irradiations were performed with the microfocus beam.

The test fixture used is the same as had been used for the previous tests. A light vacuum was pulled on the fixture after the sample was loaded. Also, we used the thinned end cap (0.14" stainless steel) as opposed to the 0.25" cap used in [1]. Figure 20 shows the flux enhancement that we achieved by thinning the end cap. The end of the Swagelok fixture was placed 1mm from the exit plane, the same as was done in [1]. Figure 21 and Figure 22 show views of the text fixture and the x-ray tube.

The user interface for North Star's machine reports two currents, the set-point current (at the electron source) and the current measured at the target. We chose a set-point current of 1.2mA to achieve a 1.0mA current at the target. Sample SL16 from [1] was irradiated with a 1mA current but does not state which current this is. Our choice of 1mA at the target either matches [1] or errs 20% high (conservative).

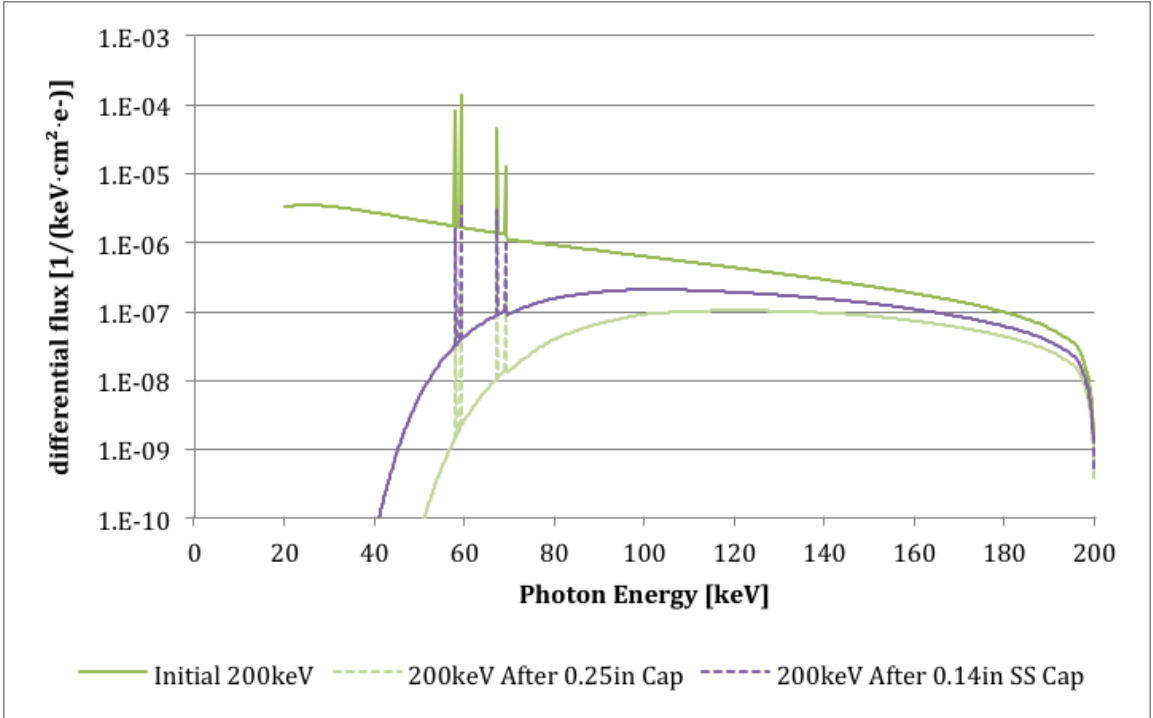
**Figure 20.** Energy spectra before and after stainless steel cap.

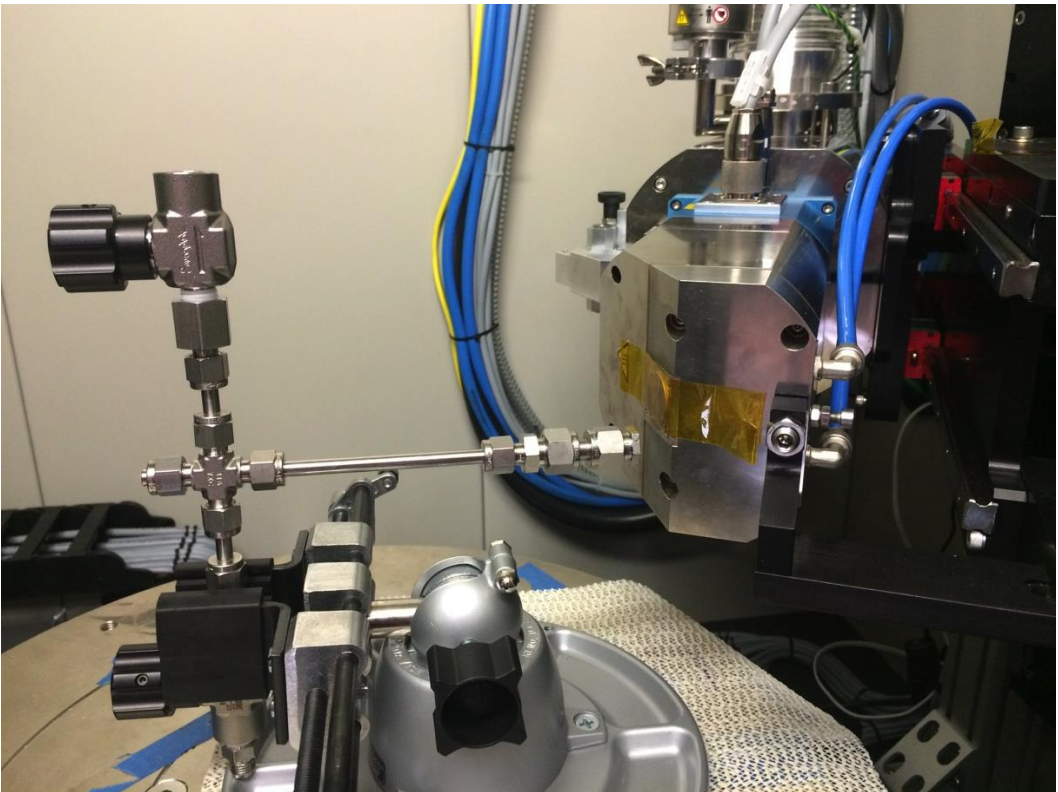
**Figure 21.** Overview of test fixture and x-ray tube.

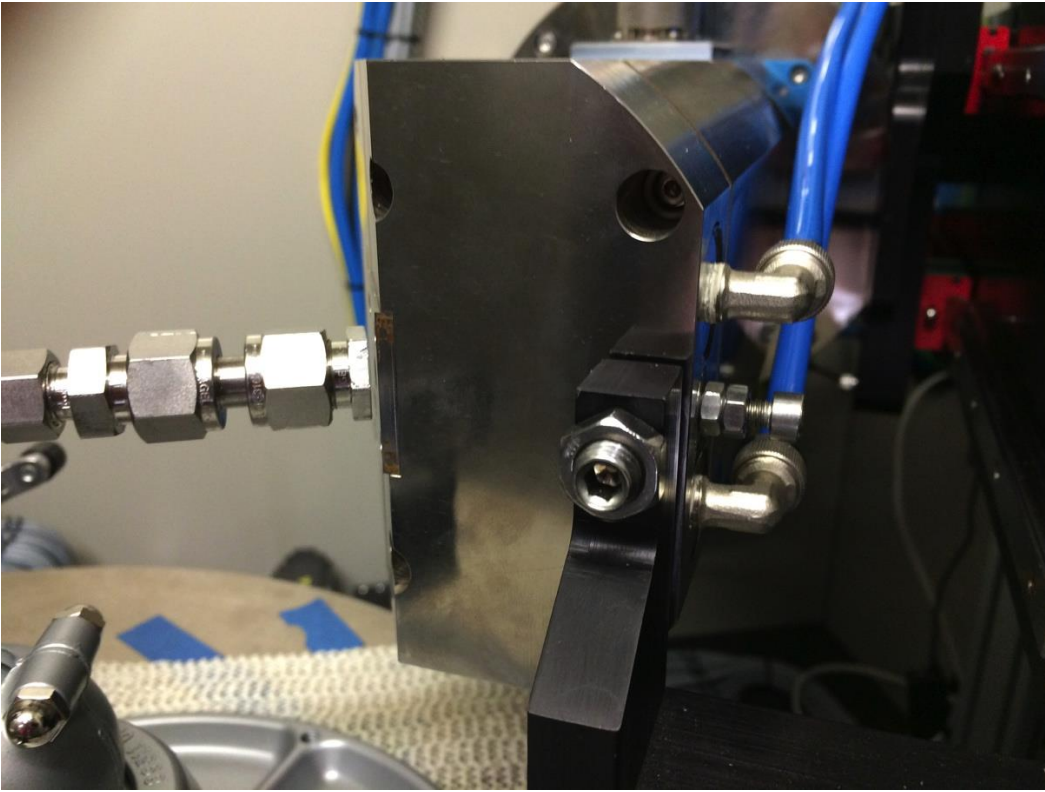
**Figure 22.** Close-up showing 1mm gap between x-ray tube and end cap of test fixture.

## 10.3. Alpha / Beta Detector

The alpha and beta measurements were conducted at JPL using the same Canberra Tennelec S5XLB system as was used previously. This detector is described in Section 2.5.   The lag time between the end of irradiation and the alpha / beta measurements was necessarily longer in the current dataset because of the distance between the x-ray facility and the on-lab detector.  The additional lag time would not prevent us from detecting of beta activity, given the lifetime of months reported in [1]. The lifetime of the alpha's reported in [1] was not accurately measured, but was clearly shorter than the beta lifetime. It is unclear whether the extra 1 hour of lag time for our 200kV samples, as compared to SL16 in [1], could explain the lack of an alpha signal.

## 10.4. Results

Results for the present work are summarized in Table 8. The measured activities (in decays per minute) are consistent with zero within uncertainties and are all below the Minimum Detectable Amount (MDA). All of these observations are consistent with no activation of the target materials. Since our run conditions were the same (max energy, beam intensity) or more severe (energy spectrum) than those in [1], our results suggest that the activity observed in [1] were not x-ray induced.

**Table 8.** Results summary with comparison to "SL16", a similar test sample from [1].

| Sample | Ref. [1] | 200kV Data Set | | | | Units |
|---|---|---|---|---|---|---|
| | SL16 | 9 | 10 | 11 | 12 | |
| Materials | Deuterated | Deuterated | Control | Deuterated | Control | |
| **Mass (TiD: DPE or TiH: PE)** | 0.594: 0.454 | 0.5: 0.5 | 0.5: 0.5 | 0.5: 0.5 | 0.5: 0.5 | g |
| **Max X-Ray Energy** | 200 | 200 | 200 | 200 | 200 | keV |
| **Shielding from end cap** | 0.25" SS | 0.14" SS | 0.14" SS | 0.14" SS | 0.14" SS | inches |
| **X-Ray Current** | 1 | 1 | 1 | 1 | 1 | mA |
| **Run Time** | 60 | 60 | 60 | 60 | 60 – 74* | minutes |
| **Lag Time (x-ray to detection)** | 30 | 103 | 92 | 96 | 94 | min |
| **Activity After X-Ray - Alpha** | 8.44±2.1 | 0.17±0.27 | -0.09±0.09 | 0.17±0.27 | 0.95±0.53 | dpm |
| **Min Detectable Activity - Alpha** | 2.54 | 1.27 | 1.27 | 1.27 | 1.27 | dpm |
| **Activity After X-Ray - Beta** | 37.95±6.47 | -1.68±0.77 | -1.89±0.74 | -1.05±0.85 | -0.63±0.90 | dpm |
| **Min Detectable Activity - Beta** | 6.93 | 3.75 | 3.75 | 3.75 | 3.75 | dpm |

*X-ray beam was found to have shut off at some time between the 11:44 and 11:59 AM monitoring points, so this 15-minute portion of the exposure was repeated in its entirety